\documentclass[paper]{elsarticle}
\graphicspath{ {./} }
\usepackage{hyperref}
\usepackage{float}
\usepackage{verbatim}
\usepackage{apalike}
\restylefloat{figure}
\restylefloat{table}
\usepackage{appendix}
\usepackage{amsmath}
\usepackage{multirow}
\usepackage{xcolor}

\journal{Expert Systems with Applications}

\biboptions{authoryear}

\begin{document}

\null 
\vfill 

\begin{center}
    \Large \textbf{Copyright Information} \\
    \vspace{1cm} 
    \normalsize
    This work is licensed under the Creative Commons Attribution-NonCommercial-NoDerivatives (CC BY-NC-ND) 4.0 International License. \\
    \vspace{0.5cm} 
    To view a copy of this license, visit \\
    \url{https://creativecommons.org/licenses/by-nc-nd/4.0/}
\end{center}

\vfill 
\newpage 

\begin{frontmatter}









\title{Predicting risk/reward ratio in financial markets for asset management using machine learning}

\author[label1,label2]{Reza Yarbakhsh \corref{cor1}}
\ead{yarbakhsh@ut.ac.ir}

\author[label2]{Mahdieh Soleymani Baghshah}
\ead{soleymani@sharif.edu}

\author[label1, label2]{Hamidreza Karimaghaie}
\ead{karimaghaie@ut.ac.ir}

\cortext[cor1]{Corresponding author}
\address[label1]{Department of Business Management, University of Tehran, Tehran, Iran}
\address[label2]{Department of Computer Science and Engineering, Sharif University of Technology, Tehran, Iran}

\begin{abstract}
Financial market forecasting remains a formidable challenge despite the surge in computational capabilities and machine learning advancements. While numerous studies have underscored the precision of computer-generated market predictions, many of these forecasts fail to yield profitable trading outcomes. This discrepancy often arises from the unpredictable nature of profit and loss ratios in the event of successful and unsuccessful predictions. In this study, we introduce a novel algorithm specifically designed for forecasting the profit and loss outcomes of trading activities. This is further augmented by an innovative approach for integrating these forecasts with previous predictions of market trends. This approach is designed for algorithmic trading, enabling traders to assess the profitability of each trade and calibrate the optimal trade size. Our findings indicate that this method significantly improves the performance of traditional trading strategies as well as algorithmic trading systems, offering a promising avenue for enhancing trading decisions.
\end{abstract}


\end{frontmatter}

\section{Introduction}
\label{introduction}

Financial markets are complex and ever-changing, rendering market forecasting a challenging task \citep{maasoumi2002entropy, raubitzek2022exploratory}. The issues arise primarily because financial markets are driven by a multitude of external factors such as political turmoil \citep{hillier2019political}, global economy \citep{matkovskyy2019financial}, investor sentiments \citep{li2014news, broadstock2019social}, and the fundamental characteristics of assets \citep{wafi2015fundamental}. While these variables are mostly unpredictable, and their information is not readily available to the public, they have a direct and strong correlation with asset prices. Despite this, with markets not being fully efficient due to information disparities \citep{barr1998persuasive}, information relevant to the market's future is reflected in its pricing \citep{dias2020random, ozkan2021impact}. Technical analysis approaches were previously used to forecast markets using historical pricing data \citep{park2007we}. Nowadays, machine learning methods have significantly improved these predictions and increased their accuracy substantially \citep{STANKOVIC201551, ayala2021technical}.
However, most machine learning algorithms still cannot provide usable trading recommendations \citep{buczynski2021review}. This is primarily because they do not consider the potential profit and loss for every trade and the possibility of incurring great losses in wrong predictions even if there are few of them, We will delve into and address this issue in greater detail in later sections.
In the following three sections of this article, we will explore three widely used models for predicting financial markets, examining their strengths and weaknesses to gain a deeper understanding of their utility in the complex world of financial forecasting.

\subsection{Regression Models for Price Forecasting in Financial Markets}
Machine learning is commonly utilized in financial market forecasting through two key methods, price and direction forecasting \citep{emerson2019trends}. When utilizing price forecasting, models perform regression on the asset's price at a specific future time based on received characteristics \citep{lu2021cnn, hu2021survey}. However, one of the main challenges with this method is the difficulty in trading with prediction results. This is largely due to the lack of probability associated with prediction success, and the unknown potential loss if the trade is unsuccessful, while for successful trades, it's important to consider both potential profit and potential loss (risk/reward ratio).

\subsection{Classification Models for Market Direction Prediction}
Predicting a market's direction, which is a classification form of machine learning, is another approach to market prediction. With this approach, the model predicts whether the price will increase or decrease at a specific time in the future in comparison to the current price \citep{yun2021prediction, dixon2017classification}. Unlike the previous method, this prediction informs us about the likelihood of a successful trade but doesn't provide details concerning the profit/loss in case of successful or failed trades \citep{ballings2015evaluating}. To address this issue, this paper proposes a new solution by leveraging both approaches.

\subsection{triple-barrier labeling}
Triple-barrier labeling is a method introduced by \cite{de2018advances} as a solution to the problems encountered in previous methods. This approach involves defining three barriers, two horizontal and one vertical. The horizontal barriers are based on price, whereas the vertical barrier is based on time and placed at a future-specific amount. The two horizontal barriers are positioned at a certain distance below and above the current asset price. When the price hits either of the horizontal barriers first, the sample can be labeled as -1 or 1, respectively. If the price hits the vertical barrier before hitting the horizontal ones, then the label 0 can be assigned to that particular sample. Alternatively, depending on the prices being higher or lower than the initial price, the label can be set as 1 or -1 for a sample hitting the vertical barrier.

\begin{figure}[htbp]
    \centering
    \includegraphics[width=\textwidth]{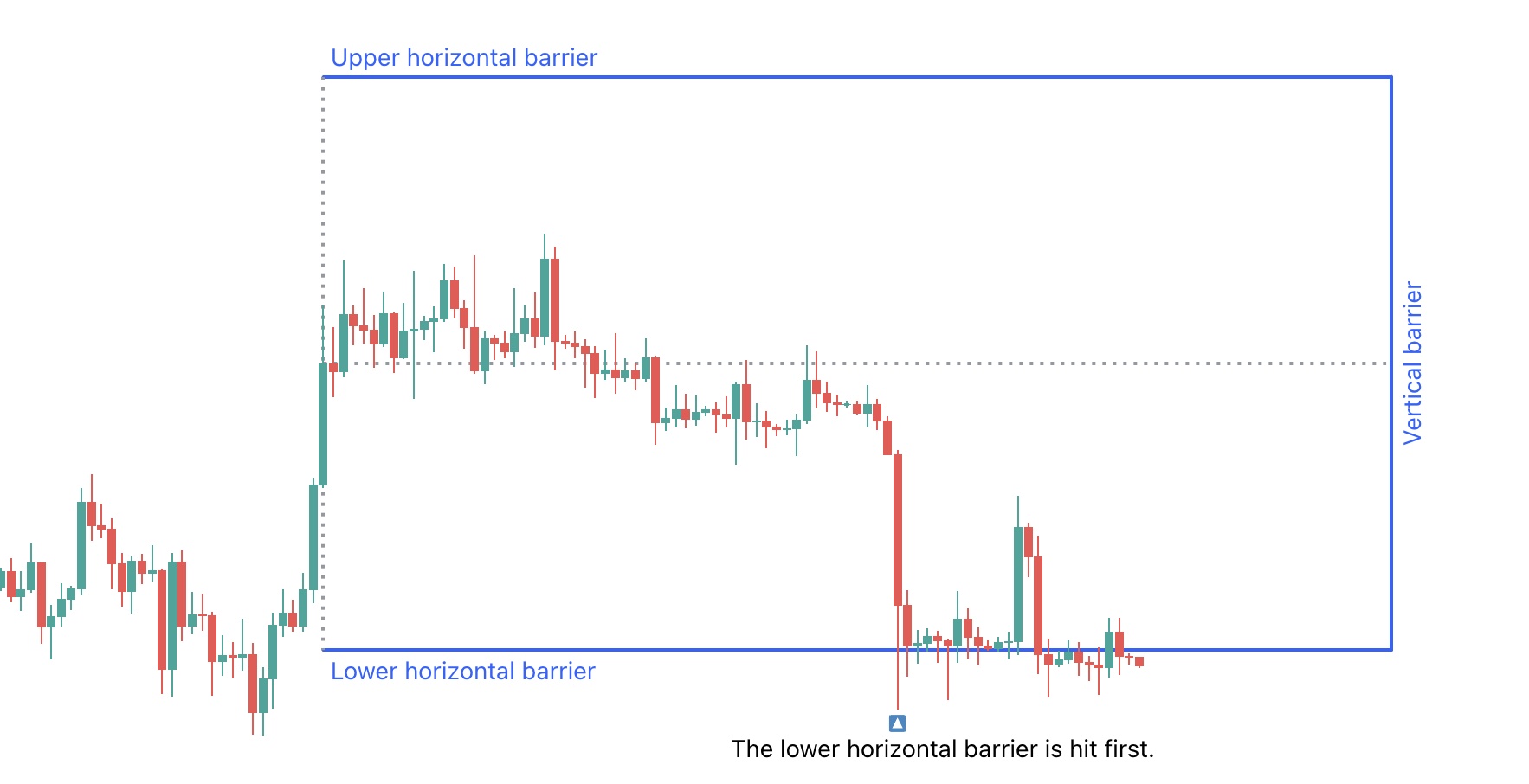}
    \caption{Triple barrier schematic}
    \label{fig:triple-barrier}
\end{figure}

If we place the take-profit and stop-loss in the ranges marked by the horizontal barriers, we can determine the amount of profit or loss in case of trading with prediction. However, as will be discussed later in the paper, the major problem with this type of labeling is the compulsory and precise determination of profit and loss limits, which results in difficulties for the model to predict accurately and train on the provided data. In addition, experimenting with different barrier distances can lead to over-fitting, which is very dangerous in financial markets \citep{de201810}.

With the growing availability of data, markets, and algorithms, financial forecasting becomes increasingly important in trading \citep{HENRIQUE2019226, gerlein2016evaluating}. Nonetheless, despite having highly accurate predictions, most forecasts are challenging to use for trade \citep{PAN201790}. Moreover, they often produce unsatisfactory trades. Hence, the usability of forecast results in automated trading is a crucial aspect to consider. In this paper, we present a novel approach that utilizes machine learning to measure every trade's potential profit or loss to enhance and optimize trading performance. The proposed technique is initially implemented within simulated environments to demonstrate its generalizability across diverse conditions. Subsequently, the strategy is applied to an empirical trend prediction model, illustrating its potential to markedly improve the efficacy of our trading activities compared to alternate trading strategies.

\section{Methodology}
\subsection{dataset}
For this study, we obtained the hourly historical OHLCV data for Bitcoin using the Binance cryptocurrency exchange. Binance was selected as the data source due to its high trading volume, which makes it a reliable and representative source of data for Bitcoin. The dataset spans from February 2017, to June 2023, and comprises five decimal columns, each of which represents a different aspect of Bitcoin's price and trading activity:

\begin{enumerate}
    \item \textbf{Open:} The opening price of Bitcoin at the beginning of the hour.
    \item \textbf{High:} The highest price reached by Bitcoin during the hour.
    \item \textbf{Low:} The lowest price reached by Bitcoin during the hour.
    \item \textbf{Close:} The closing price of Bitcoin at the end of the hour.
    \item \textbf{Volume:} The total amount of Bitcoin traded during the hour.

\subsection{Preprocessing}

\subsubsection{Feature Calculation and Transformation}
For each hourly data point, we computed an array of technical indicators across different periods. The raw OHLCV data is transformed using various technical indicators, details of which have been elaborated in the appendix.

\begin{enumerate}
    \item \textbf{Feature Normalization:} After the calculation of these indicators, data normalization was applied to each feature using the StandardScaler. This ensures a consistent influence on the model across all features and facilitates model convergence during training.
    
    \item \textbf{Data Truncation:} The initial segments of the dataset were discarded due to the unavailability or potential inaccuracy of certain indicators. This ensures the model is not influenced by potentially misleading data.
\end{enumerate}

\subsubsection{Label Construction}
Labels are devised in two distinct formats:

\begin{enumerate}
    \item \textbf{Direction Prediction:} A binary label indicating whether the price changes are positive, representing an upward price movement, or negative, indicating a downward movement and is intended for the direction prediction model.
    
    \item \textbf{Price Prediction:} This label depicts the actual price change and is intended for the price prediction model.
\end{enumerate}

\subsubsection{Weight Assignment}
Weights are designated to each data point based on the price change magnitude of Bitcoin at the respective timestamp. This approach ensures that the model gives appropriate importance to data points with significant price changes, acknowledging the inherent challenges associated with forecasting larger price movements.

\begin{table}[htbp]
    \centering
    \caption{Features}
    \label{tab:features}
    \begin{tabular}{|l|p{9cm}|p{1.3cm}|}
        \hline
        \textbf{Feature} & \textbf{Description} & \textbf{Type} \\
        \hline
        TRIX & A momentum oscillator that displays the percent rate of change of a triple exponentially smoothed moving average. & Float \\
        MACD & A trend-following momentum indicator that shows the relationship between two moving averages of a security’s price. & Float \\
        PPO & A momentum oscillator that measures the difference between two moving averages as a percentage of the larger moving average. & Float \\
        ROC & Measures the percentage change between the most recent price and a price "n" periods in the past. & Float \\
        EFI & The Ease of Movement Indicator is a volume-based oscillator that fluctuates above and below the zero line. & Float \\
        CMO & Chande Momentum Oscillator, is created by calculating the difference between the sum of all recent gains and the sum of all recent losses divided by the sum of all price movement over the period. & Float \\
        RSI & Relative Strength Index, a momentum oscillator that measures the speed and change of price movements. & Float \\
        CCI & Commodity Channel Index, a versatile indicator that can be used to identify a new trend or warn of extreme conditions. & Float \\
        WilliamsR & A momentum indicator that measures overbought and oversold levels. & Float \\
        CMF & Chaikin Money Flow measures the amount of Money Flow Volume over a specific period. & Float \\
        Price Change & Percentage-based variations in price observed over recent 5-hour sequences (price model only). & Float \\
        Market Direction & Market direction in the upcoming 5 hours. 1 for upward and -1 for downward (price model only). & Binary \\
        \hline
    \end{tabular}
\end{table}

\subsubsection{Dataset Division}

We trained our machine learning model on Bitcoin's hourly OHLCV (Open, High, Low, Close, Volume) data comprising approximately 50,000 hourly candles. The dataset was divided into three distinct sets these sets is illustrated in Figure~\ref{fig:DatasetDivision}:
\begin{itemize}
    \item \textbf{Training set}: Spanning from September 8, 2017, to August 17, 2019.
    \item \textbf{Validation set}: Beginning on August 17, 2019, and ending on January 31, 2020.
    \item \textbf{Test set}: Beginning on January 31, 2020, and ending on June 2, 2023.
\end{itemize}

\begin{figure}[htbp]
    \centering
    \includegraphics[width=\textwidth]{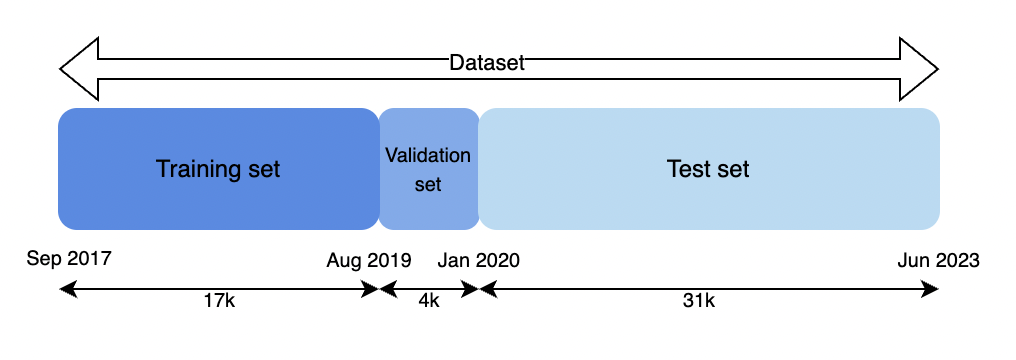}
    \caption{Dataset Division}
    \label{fig:DatasetDivision}
\end{figure}

Using a larger test set to evaluate the model's performance in diverse situations and over a long-term period offers several advantages:

\begin{enumerate}
    \item \textbf{Comprehensive Evaluation}: Allows for a more in-depth assessment over various market phases.
    \item \textbf{Real-World Scenarios}: The model encounters a broader range of real-world events, giving insight into its generalization capabilities.
    \item \textbf{Long-Term Performance}: Provides an indication of the model's long-term predictive abilities.
    \item \textbf{Identify Drift}: Easier recognition of "concept drift" where the statistical properties of the target variable change over time \citep{harries1995detecting}.
    \item \textbf{Robustness Assessment}: Evaluating the model over an extended period allows for a rigorous examination of its stability and reliability amidst fluctuating market dynamics. This long-term analysis helps to discern whether the model can consistently apply its predictive logic through various economic cycles, including growth, recession, or stability, and events that disrupt the norm, such as the volatility observed during the COVID-19 pandemic or market fluctuations in response to pronouncements by influential individuals like Elon Musk \citep{ante2023elon}.
\end{enumerate}

\subsection{Simulation of Directional Predictions}

The proposed risk management methodology's adaptability and performance are crucial to gauge across varied conditions. To achieve this, we've simulated trades influenced by synthetic directional probabilities. Trading results are then analyzed both in the presence and absence of the risk management strategy. We delineate each synthetic probability paradigm along with the underlying scientific justification:

\begin{enumerate}
\item \textbf{Balanced Probabilistic Model with a Targeted Success Rate:}
\begin{itemize}
\item \textit{Configuration:} In this model, we simulate a traditional trading strategy that is known to have a 60\% success rate. In order to achieve this, we randomly assign upward and downward predictions, so that 60\% of them are correct. Furthermore, we assign the probability of 0.6 to each prediction in our calculations for bet sizing.

\item \textit{Rationale:} The purpose of this probabilistic configuration is to assess the advantage of supplementing a conventional trading strategy with a price prediction model. By applying a static probability of 0.6 across all trades within Kelly's formula, we mimic the trader’s historical confidence level in their strategy. Our investigation seeks to demonstrate that even when a traditional strategy is coupled with a constant probability figure, the use of a price prediction model for managing risk and reward can potentially enhance overall trading performance. The approach validates the synergy between established trading strategies and advanced predictive modeling, potentially leading to more sophisticated and informed trading decisions.
\end{itemize}

\item \textbf{Optimal Predictive Model:}
\begin{itemize}
\item \textit{Configuration:} The probabilities are fixed at 0.8 for long positions and 0.2 for short ones. Additionally, predictions are consistently accurate, ensuring the model's infallibility.
\item \textit{Rationale:} Through this model, we intend to assess the upper boundary of our risk management strategy's performance, especially when supplementing near-perfect trades. It provides insights into potential return amplification in optimal conditions. Although in reality, no model can be truly infallible, especially in the volatile world of finance, this simulation showcases the potential of the proposed strategy to increase the profitability and Sharpe Ratio, even for an idealistic trading model with no incorrect predictions.
\end{itemize}

\item \textbf{Probabilistic Model Based on Gaussian Distribution:}
\begin{itemize}
\item \textit{Configuration:} Probabilities are generated using a Gaussian distribution with means of 0.6 for long positions and 0.4 for short ones. The target success rate is maintained at 60\%.
\item \textit{Rationale:} Emulating real-world trend prediction mechanisms, this model offers a realistic evaluation spectrum. It serves to empirically validate the risk management technique's resilience and adaptability amidst typical market conditions.
\end{itemize}
\end{enumerate}

By methodically analyzing trade outcomes across these varied synthetic probability models, we aim to accentuate the comprehensive applicability and robustness of our risk management approach. This further substantiates its merits as an augmentative tool for diverse trading strategies.

\end{enumerate}
\subsection{direction prediction model}
A predictive model was developed and implemented utilizing XGBoost to ascertain the trajectory of the Bitcoin market. XGBoost was selected as an optimal model choice for this investigation due to its efficacy in detecting non-linear associations and mitigating overfitting concerns in small datasets \citep{raubitzek2022exploratory,moller2016photometric}. Given the predominantly non-linear nature of financial data relationships \citep{abhyankar1997uncovering,mcmillan2003non} and the scarcity of training data, particularly in light of the adaptive nature of financial markets where older data hold limited relevance for learning \citep{lo2017adaptive} XGBoost was deemed a fitting choice for this study.

In developing our model, we incorporated an array of technical indicators as features. We strategically selected these indicators based on their low inter-correlation to optimize the accuracy and robustness of the model \citep{hall1999correlation}. The indicators were computed across various periods and parameters, culminating in a compendium of 550 unique features for each sample.

The label assigned to each sample represents whether the Bitcoin price will be higher (labeled 1) or lower (labeled -1) than the current price in the next five hours. The weight assigned to each sample is based on the magnitude of this difference, enabling the model to predict larger profits and losses more accurately.

\subsection{price change prediction model}
With knowledge of the probability of price movement direction in the next five hours, the next step towards executing profitable trades involves estimating potential profits and losses. To achieve this, we trained a regressor model instead of a binary classifier, and used the price change percentage in the next five hours as the target.

The price change prediction model utilizes the same features as the direction prediction model, while incorporating two additional features: the price change percentage over recent 5-hour intervals and the market direction for the upcoming 5 hours.

The first feature delves into the percentage-based variations in price observed over recent 5-hour sequences (from $i-5$ to $i$). This feature offers a deeper insight into the movement magnitudes, which is pivotal for making precise future price change predictions.

The second feature, pinpointing the actual direction the market takes in the subsequent 5-hour span, is particularly instrumental. By offering insights into the future direction of the market, this feature equips the price change prediction model with the capacity to anticipate price fluctuations in both bullish and bearish market scenarios. This feature is of high importance because it gives the price change prediction model insights about the future direction of the market which would be used later in our trading strategy to predict the price change in both upward market and downward market scenario and use the results as profit and loss estimates.

\begin{figure}[htbp]
    \centering
    \includegraphics[width=\textwidth]{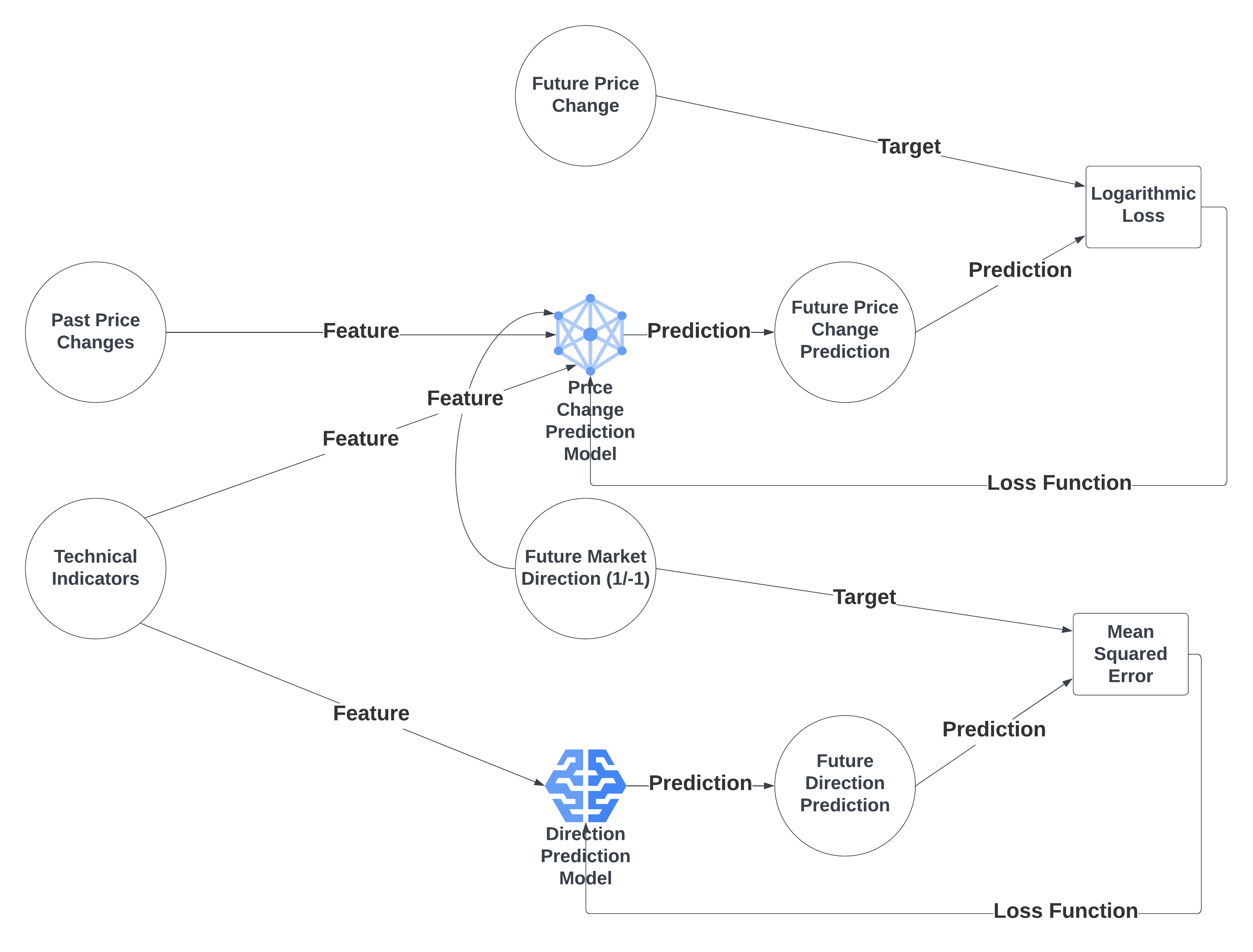}
    \caption{Training Process}
    \label{fig:training}
\end{figure}

It's important to note that this methodology doesn't involve the inappropriate use of future data. Instead, it provides a systematic way to make informed trading decisions by accounting for both potential market directions. The training process of our direction prediction and price change models can be seen in Figure~\ref{fig:training}, while the interconnected trading process using these models is illustrated in Figure~\ref{fig:trading}.

For trading, we generate two separate predictions for each trading scenario in our price change model. The first prediction assumes a bullish market direction, while the second one assumes a bearish market direction. By doing this, we can produce estimates for price changes in both upward and downward market scenarios.

To align these predictions with actual trading decisions, we reference our direction prediction model. If this model forecasts an upward market movement, we treat the bullish price change estimate as potential profit and the bearish estimate as potential loss. Conversely, if a downward market is predicted, the bullish estimate becomes our potential loss, and the bearish estimate our potential profit. As an illustration, supposing the direction model prognosticates a market uptrend with a likelihood of 60\%. In our proprietary price forecasting model, we presuppose the market's direction to be affirmative or, alternatively, adverse. In the event of an affirmative market trend, our price projection entails a 5\% increment, while in the case of an adverse market trend, it entails a 4\% decrement. Employing this aforementioned information, we can refine our risk appraisal, subsequently leveraging this data to implement risk management strategies with greater precision.

\subsection{trading strategy}
With the probability of upward and downward market movements established from our direction prediction model, and a dependable estimate of the magnitude of price changes in each direction from our price change prediction model, we are better positioned to gauge potential profits and losses from trading. This enables us to make well-informed decisions about the direction and magnitude of a trade.

To determine the optimal size of our trade, we turn to Kelly criterion, a well-regarded formula long associated with successful trading strategies in financial markets \citep{thorp2008kelly}. The Kelly formula helps traders ascertain the most appropriate size for a trade based on the following equation:

$$
f^{*}=\frac{p}{a}-\frac{q}{b}
$$

The optimal trade size relative to total assets is denoted by $f^{*}$. The likelihood of a market trending upwards is indicated by $p$, while $q$ symbolizes the odds of a downward market shift. The price alterations during an upward and downward market trend are represented by $a$ and $b$ respectively. The resulting value obtained from the Kelly formula dictates not only the direction but also the size of the trade. It is worth noting that the resulting trade size is optimized for long-term profit and does not take into account risk considerations \citep{maclean2010good}. Various solutions exist to address the problem of risk management, although they are beyond the scope of this paper \citep{baker2013optimal}.

\begin{figure}[htbp]
    \centering
    \includegraphics[width=\textwidth]{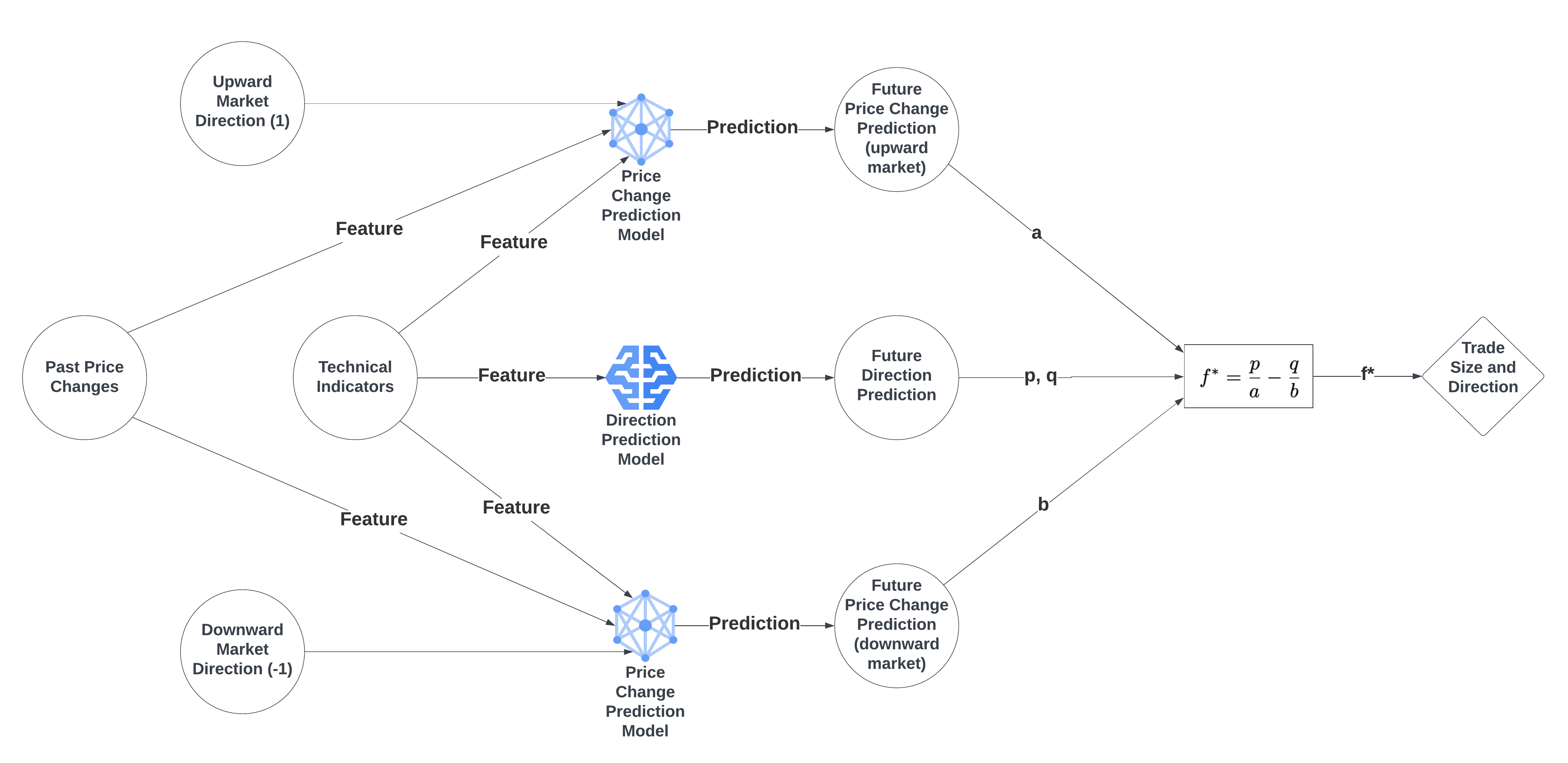}
    \caption{Trading Process}
    \label{fig:trading}
\end{figure}

The predictions generated by our direction prediction model offer insight into the market's direction, with $p$ indicating the probability of an upward movement and $q=1-p$ showing downward movement probability. On the other hand, our price change prediction model predicts the price change amount associated with both upward and downward directions. Specifically, $a$ reflects the predicted upward price change, while $b$ denotes the predicted downward price change by the second model. The trading process using Kelly criterion and both our models is depicted in Figure~\ref{fig:trading}.

\subsection{Bet Sizing Based on Predicted Probabilities}

\subsubsection{Kelly Criterion}

Let's derive the Kelly Criterion by maximizing the expected logarithmic growth of the bankroll.

\textbf{Assumptions:}
\begin{enumerate}
    \item A bettor wants to bet a fraction \( f \) of their current bankroll.
    \item With probability \( p \), the bettor wins the bet, and their bankroll is multiplied by \( 1 + af \) (since they receive \( a \) times their bet in profit plus their original bet).
    \item With probability \( q \) (where \( q = 1-p \)), the bettor loses the bet, and their bankroll is multiplied by \( 1 - bf \) (since they lose their bet).
\end{enumerate}

The bettor aims to maximize the expected value of the logarithm of the resulting bankroll. This is given by:
\[
E[\ln(\text{Bankroll})] = p \ln(1 + af) + q \ln(1 - bf)
\]

To find the maximum, differentiate with respect to \( f \) and set the result equal to zero:
\[
\frac{d}{df} \left( p \ln(1 + af) + q \ln(1 - bf) \right) = p \frac{a}{1+af} - q \frac{b}{1-bf} = 0
\]

Solving for \( f \), we get the Kelly fraction:
\[
f^* = \frac{p}{a} - \frac{q}{b}
\]

This formula gives the optimal fraction \( f^* \) of the current bankroll to bet in order to maximize the expected logarithmic growth of the bankroll.

Given a model that predicts the probability \( p \) of the market moving upward, the Kelly Criterion can be applied to determine the optimal fraction of your portfolio to invest. The Kelly fraction is given by:

\[
f^* = \frac{p}{a} - \frac{1-p}{b}
\]

Where:
\begin{itemize}
    \item \( p \) is the probability the market moves upward.
    \item \( 1-p \) is the probability the market moves downward.
    \item \( a \) represents the net fractional gain when the market moves up.
    \item \( b \) represents the net fractional loss when the market moves down.
\end{itemize}

\textbf{Interpreting the Results:}
\begin{enumerate}[i)]
    \item \textbf{Positive \( f^* \)}: Indicates a long position, betting a fraction \( f^* \) of your bankroll or portfolio on the market moving upwards.
    \item \textbf{Negative \( f^* \)}: Suggests a short position, with the absolute value of \( f^* \) giving the optimal fraction of your bankroll to short.
    \item \textbf{\( f^* > 1 \) or \( f^* < -1 \)}: Indicates that the optimal bet size exceeds your current bankroll, suggesting the use of leverage in your trades.
\end{enumerate}

\textbf{Considerations:}
\begin{enumerate}
    \item \textbf{Leverage}: In the context of trading, \textit{leverage} refers to the ability to control a large position with a relatively small amount of capital. Leverage is typically expressed as a ratio, such as 2:1 or 10:1, which means that for every dollar of your own capital, you can trade with 2 or 10 dollars respectively.
    \item \textbf{Risk Management}: The Kelly Criterion maximizes logarithmic growth but doesn't account for all types of risk or individual risk tolerances. Ensure other risk management strategies are in place.
    \item \textbf{Model Accuracy}: The efficacy of the Kelly Criterion is contingent on accurate probability predictions. Regularly validate and test the model's performance.
    \item \textbf{Fractional Kelly}: Many traders use a fraction of the Kelly recommendation (e.g., half-Kelly) for a more conservative approach, especially when estimated probabilities are uncertain.
\end{enumerate}

\textbf{Visualizing the Kelly Criterion:}

\begin{enumerate}
    \item \textbf{Kelly Criterion 3D Plot} \\
    In this plot, the Kelly Criterion, often used in gambling and investing to determine the optimal bet size, is visualized in a three-dimensional space. The x-axis represents the probability of a win $p$, while the y-axis (on a logarithmic scale) represents the proportion of the bet $b$. The z-axis gives us the fraction of the bankroll $f$ that should be staked based on the Kelly formula. The colormap `viridis' provides a visual representation of the varying values of $f$, helping us understand the relationship between $p$, $b$, and $f$.
    
\begin{figure}[htbp]
    \centering
    \includegraphics[width=\textwidth]{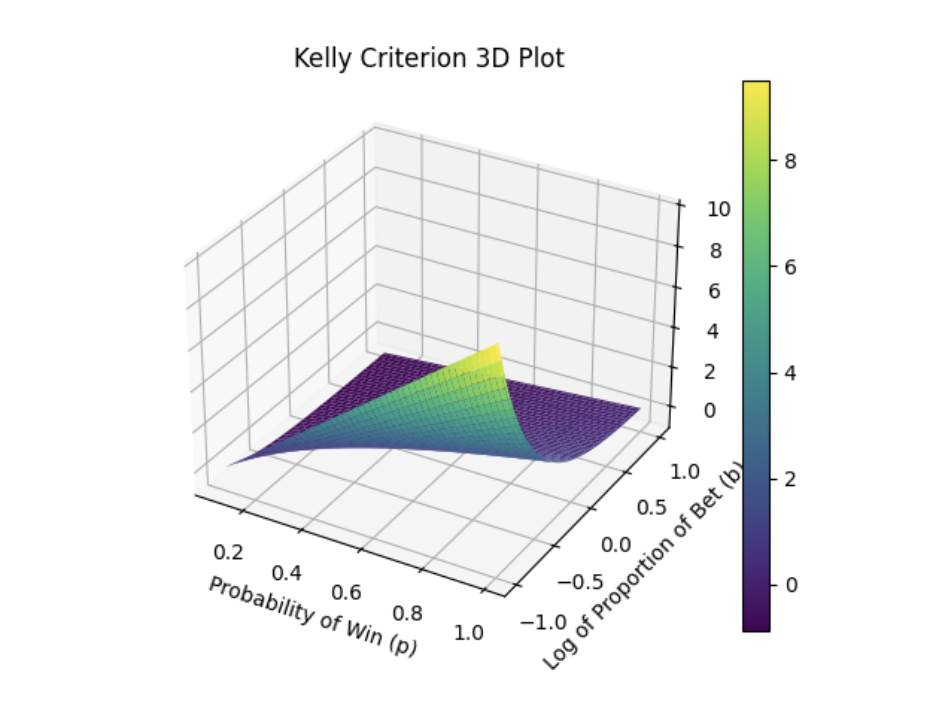}
    \caption{Kelly Criterion visualization 1}
    \label{fig:kelly1}
\end{figure}
    
    \item \textbf{Kelly Fraction across p and a=b} \\
    Here, we visualize the Kelly Fraction across varying probabilities $p$ and a condition where both $a$ and $b$ are the same, represented as $a=b$. The x-axis showcases the range of probabilities from 0 to 1, and the y-axis provides values of $a=b$ from 0.1 to 1. The z-axis, labeled $f^*$, displays the Kelly fraction. The surface plot offers insights into how the Kelly fraction changes with $p$ and $a=b$, using the `viridis' colormap for added clarity.

\begin{figure}[htbp]
    \centering
    \includegraphics[width=\textwidth]{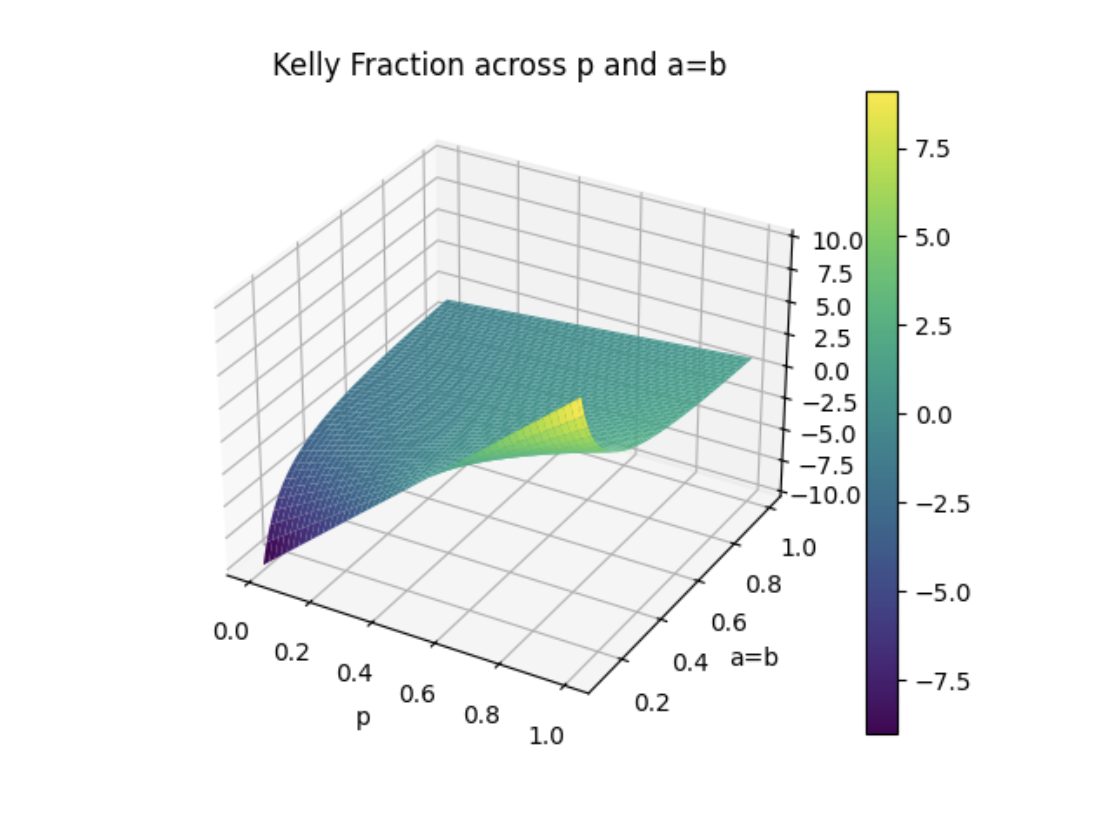}
    \caption{Kelly Criterion visualization 2}
    \label{fig:kelly2}
\end{figure}
    
    \item \textbf{Kelly Fraction for p=0.6 over varying a and b} \\
    This plot captures the Kelly Fraction over a spectrum of $a$ and $b$ values, specifically for $p=0.6$. The x and y axes represent the ranges of $a$ and $b$ values, respectively, while the z-axis displays the calculated Kelly Fraction $f^*$. With the `viridis' colormap in place, this 3D visualization provides an understanding of how the fraction $f^*$ varies when adjusting $a$ and $b$, keeping $p$ constant at 0.6.

\begin{figure}[htbp]
    \centering
    \includegraphics[width=\textwidth]{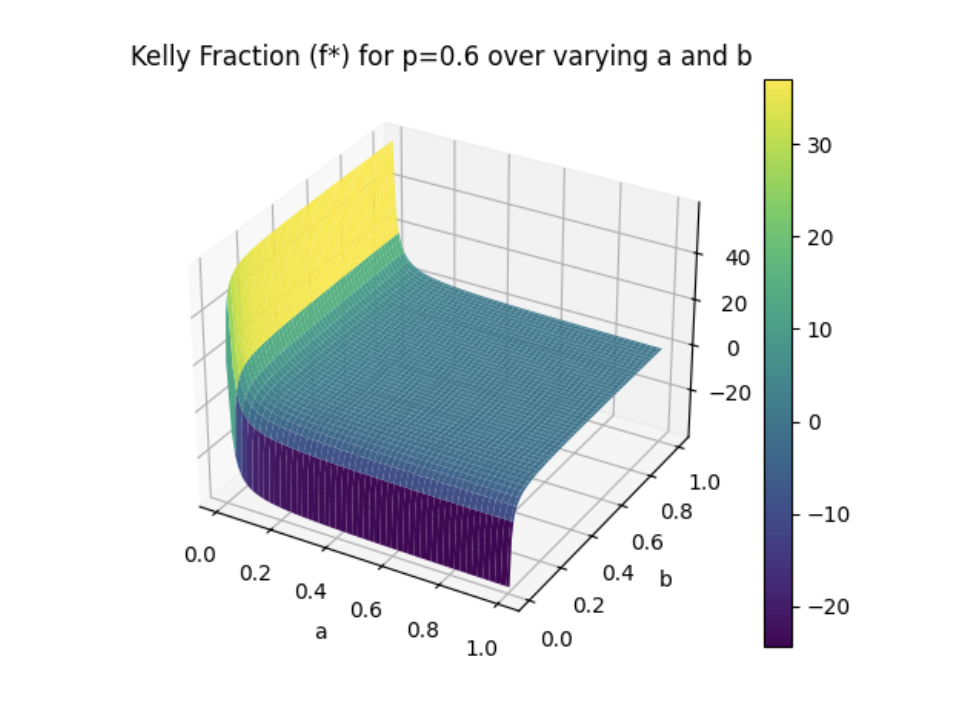}
    \caption{Kelly Criterion visualization 3}
    \label{fig:kelly 3}
\end{figure}

\end{enumerate}

\subsubsection{Gaussian RM}

The bet sizing approach detailed here is adapted from Dr. Marcos López de Prado's book "Advances in Financial Machine Learning". This approach calculates the optimal bet size using predicted probabilities. The bet size is determined based on how much the predicted probability deviates from an expected value, typically set at 0.5 (representing a 50\% chance) \citep{de201810}.

\textbf{Formula:}
If the predicted probability (\texttt{prob}) is less than or equal to the expected probability, the bet size is set to 0. Otherwise, the bet size \( m \) is given by:
\begin{equation}
m = 2Z[z] - 1
\end{equation}
Where \( Z \) is the standard normal cumulative distribution function (CDF) and \( z \) is the test statistic:
\begin{equation}
z = \frac{\texttt{prob} - \texttt{expected}}{\sqrt{\texttt{prob}(1-\texttt{prob})}}
\end{equation}

In essence, the function uses the difference between the predicted and expected probabilities to determine the bet's size, adjusting the size according to the confidence level represented by the predicted probability.
Details and results of this bet sizing approach will be provided in the subsequent sections of this article. We will further compare this approach with the Kelly criterion to offer a comprehensive understanding of its efficacy.

\section{results}

\subsection{Introduction to Results}
The aim of this section is to provide a detailed presentation and analysis of the results obtained from our study. These results not only validate our methodologies but also underline the significance of our contributions in the broader context of financial risk management and trading strategy optimization.

\subsection{Price Change Prediction}

To determine the effectiveness of the price change prediction model—serving as the foundation for risk and reward parameters in our strategies—we employed a set of evaluation metrics. These quantitative metrics elucidate the model's proficiency in predicting price alterations, a critical component for our overarching trading approach.

\begin{table}[h!]
\centering
\begin{tabular}{|l|c|}
\hline
\textbf{Metric} & \textbf{Value} \\
\hline
Mean Absolute Error (MAE) & 0.0079 \\
\hline
Mean Squared Error (MSE) & 0.00015 \\
\hline
Root Mean Squared Error (RMSE) & 0.012 \\
\hline
R-squared & 0.12 \\
\hline
\end{tabular}
\caption{Evaluation metrics for the price change prediction model}
\end{table}

It is crucial to note that the results presented pertain solely to predictions where the "future direction" feature aligns with the actual market direction. Evaluating the outcomes where the market behaved inversely in the past is infeasible. Nevertheless, the trading outcomes, based on the reward-to-loss ratio from this model's predictions, indicate its commendable ability to discern potential rewards and losses.

\subsection{Synthetic Prediction Results}

\subsubsection{Context and Methodology}
The synthetic directional probabilities serve as a robust mechanism to simulate trading under various probabilistic models. This study focuses on four core performance metrics—Return, Drawdown, Sharpe Ratio, and RoMaD (Return Over Maximum Drawdown)—to evaluate trading performance under these synthesized scenarios. A constant modifier was uniformly applied across all positions in each strategy to streamline comparisons. It's pertinent to mention that this application has negligible effects on the Sharpe Ratio, as elaborated by \citep{gordon2003var}.

\subsubsection{Performance Metrics Analysis}
The encompassing results of our trade simulations across the different probabilistic models are captured in Table \ref{table:performance}. This table offers a holistic view of how the proposed risk management methodology fares against varying synthetic directional probabilities.

\begin{table}[h!]
\centering
\begin{tabular}{|l|c|c|c|c|}
\hline
\textbf{Model} & \textbf{Metric} & \textbf{Without RM} & \textbf{Gaussian RM} & \textbf{Kelly RM} \\
\hline
\multirow{4}{*}{Gaussian Model} 
& Return & 41.51\% & 74.74\% & 84.07\% \\
& Drawdown & 0.27\% & 0.67\% & 0.60\% \\
& RoMaD & 40.35 & 27.40 & 33.41 \\
& \textcolor{blue}{Sharpe Ratio} & \textcolor{blue}{6.81} & \textcolor{blue}{9.02} & \textcolor{blue}{9.22} \\
\hline
\multirow{4}{*}{Balanced Model} 
& Return & 49.93\% & 67.06\% & 78.04\% \\
& Drawdown & 0.44\% & 0.43\% & 0.31\% \\
& RoMaD & 29.38 & 38.95 & 61.69 \\
& \textcolor{blue}{Sharpe Ratio} & \textcolor{blue}{7.47} & \textcolor{blue}{11.30} & \textcolor{blue}{12.81} \\
\hline
\multirow{4}{*}{Optimal Model} 
& Return & 677.77\% & 804.44\% & 1315.61\% \\
& Drawdown & 0.00\% & 0.00\% & 0.00\% \\
& RoMaD & N/A & N/A & N/A \\
& \textcolor{blue}{Sharpe Ratio} & \textcolor{blue}{8.22} & \textcolor{blue}{53.87} & \textcolor{blue}{94.53} \\
\hline
\end{tabular}
\caption{Simulation results for different risk management strategies}
\label{table:performance}
\end{table}

\subsubsection{Graphical Representation}
To offer a visual comparison and deeper insight into risk-adjusted performance, we have charted the Sharpe Ratios of different risk management strategies.

\begin{figure}[h!]
\centering
\includegraphics[width=0.8\textwidth]{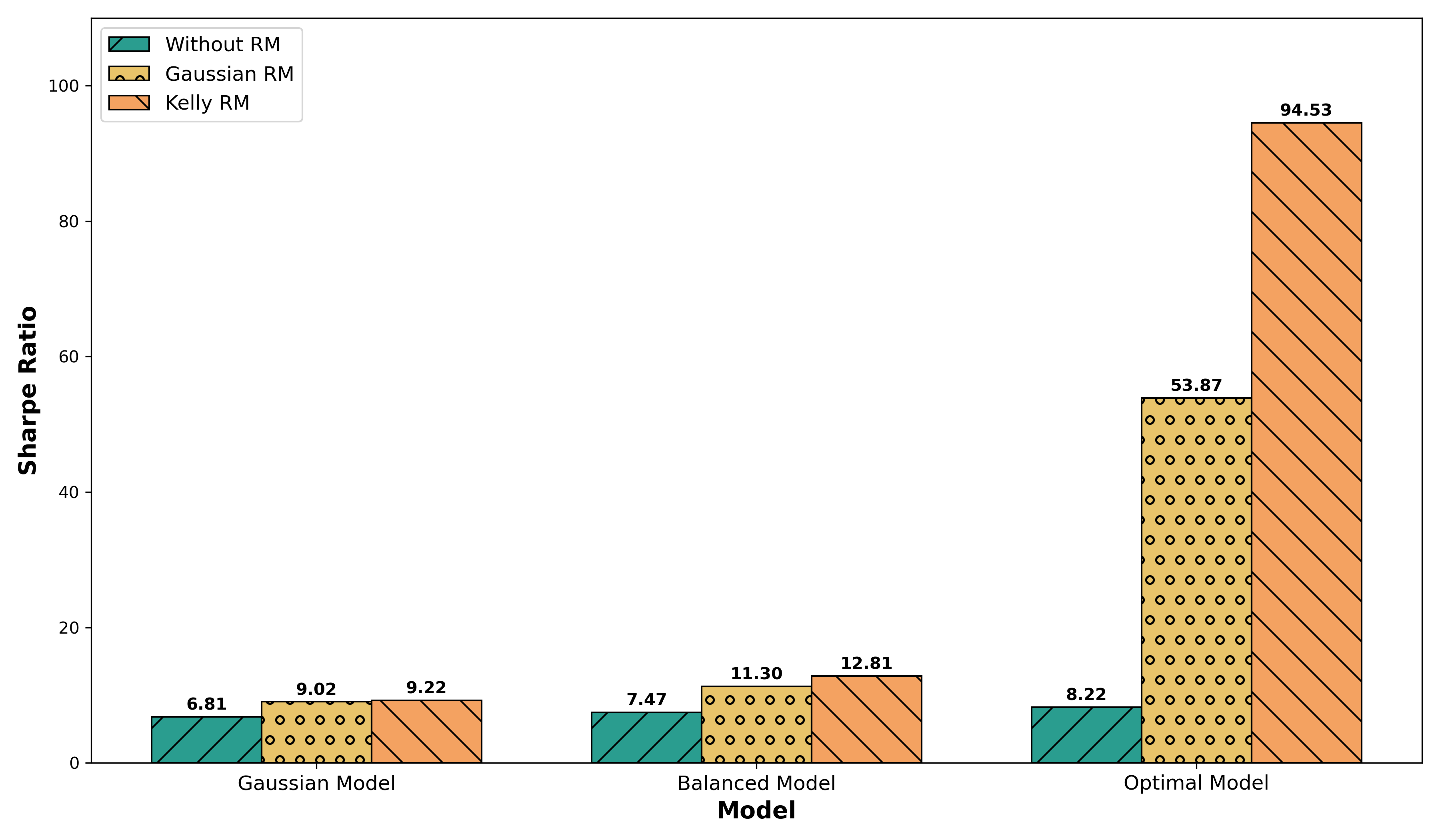}
\caption{Sharpe Ratio for different risk management strategies}
\end{figure}

\subsubsection{Summary of Synthetic Prediction}
Our synthetic predictions underscore the criticality of incorporating an efficient risk management strategy. Among the findings, the significant enhancement in the Sharpe ratio, particularly when applying risk management, stands out. It highlights the proposed methodology's prowess in achieving superior risk-adjusted returns across a gamut of probabilistic scenarios.

\subsection{Direction Prediction}

\subsubsection{Backtesting Context and Methodology}
Our research utilized the Bitcoin dataset for evaluating the proposed trading strategy. This dataset was partitioned into three phases. The training phase extended from September 2017 to August 2019, incorporating 17,000 data points. Following this, the validation phase ran up until February 2020, including an additional 4,000 data points. Lastly, the testing phase spanned from February 2020 to June 2023, embracing a comprehensive 29,000 data points.

\subsubsection{Benchmark Strategies}
To ascertain the effectiveness of our proposed strategy, it was juxtaposed against three distinct trading methodologies. The basic "buy and hold" strategy serves as a passive investment approach. The triple barrier labeling method, on the other hand, strategizes exits based on predetermined price movement barriers. Lastly, the side learning method capitalizes on market direction predictions without considering risk-reward ratios.

\subsubsection{Model Training Metrics}
For an in-depth evaluation of the model during its training phase, various metrics were employed, including accuracy, F1 score, the confusion matrix, and the precision-recall curve. These metrics furnish a comprehensive view of the model's prediction capabilities and general performance.

\begin{table}[h!]
\centering
\begin{tabular}{|l|l|c|c|c|c|}
\hline
\textbf{Model} & \textbf{Metric} & \textbf{Logloss} & \textbf{Precision} & \textbf{Recall} & \textbf{F1-score} \\
\hline
\multirow{5}{*}{\textbf{Triple Barrier}} & Overall & 0.6913 & - & - & 0.53 \\
& Downtrend (0) & - & 0.53 & 0.49 & 0.51 \\
& Upward Trend (1) & - & 0.53 & 0.57 & 0.55 \\
& Macro Avg & - & 0.53 & 0.53 & 0.53 \\
& Weighted Avg & - & 0.53 & 0.53 & 0.53 \\
\hline
\multirow{5}{*}{\textbf{Side Learning}} & Overall & 0.6892 & - & - & 0.54 \\
& Downward Trend (0) & - & 0.53 & 0.44 & 0.48 \\
& Uptrend (1) & - & 0.54 & 0.63 & 0.58 \\
& Macro Avg & - & 0.54 & 0.54 & 0.53 \\
& Weighted Avg & - & 0.54 & 0.54 & 0.53 \\
\hline
\end{tabular}
\caption{Training Metrics for Trend Prediction and Triple Barrier Models}
\label{table:training_metrics}
\end{table}

\begin{figure}[h!]
\centering

\begin{minipage}{0.48\textwidth}
\centering
\includegraphics[width=0.98\textwidth]{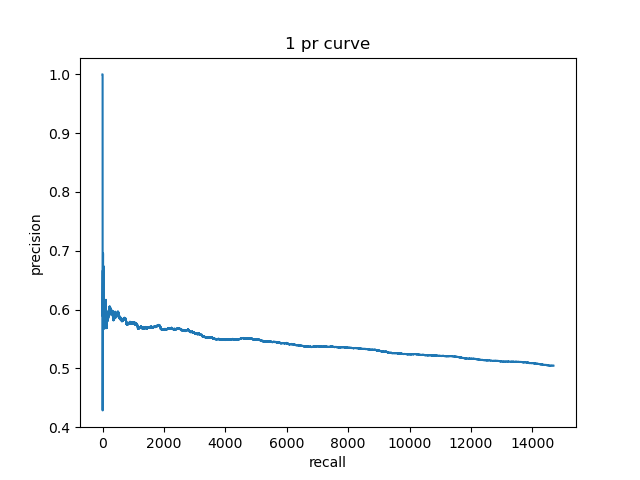}
\end{minipage}
\hspace{0.02\textwidth}
\begin{minipage}{0.48\textwidth}
\centering
\includegraphics[width=0.98\textwidth]{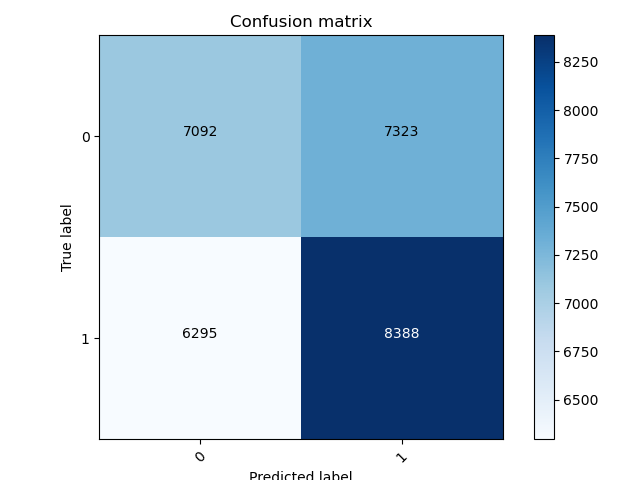}
\end{minipage}

\caption{Precision-Recall Curve and Confusion Matrix for Triple Barrier Model}
\label{fig:barrier_metrics}
\end{figure}

\begin{figure}[h!]
\centering

\begin{minipage}{0.48\textwidth}
\centering
\includegraphics[width=0.98\textwidth]{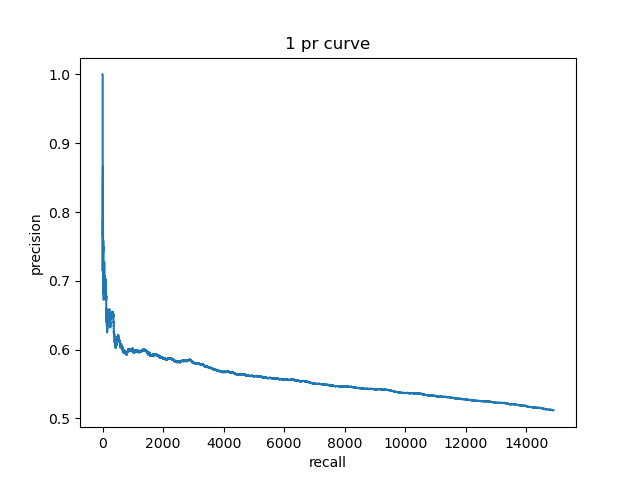}
\end{minipage}
\hspace{0.02\textwidth}
\begin{minipage}{0.48\textwidth}
\centering
\includegraphics[width=0.98\textwidth]{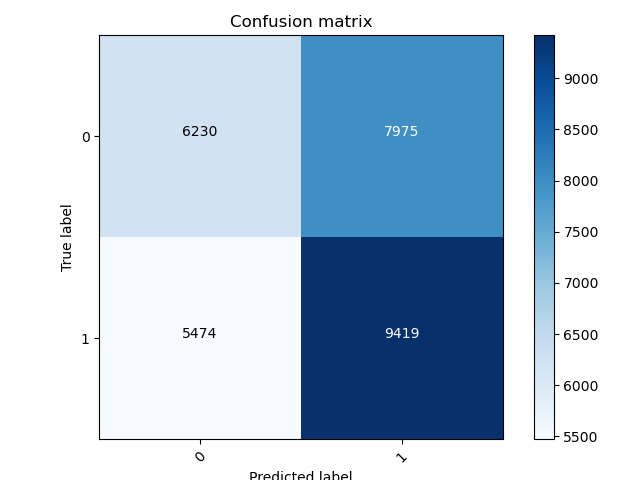}
\end{minipage}

\caption{Precision-Recall Curve and Confusion Matrix for Side Learning Model}
\label{fig:trend_metrics}
\end{figure}

\subsubsection{Performance Metrics for Backtesting}
For the backtesting, the chosen performance metrics encompassed Cumulative Return, Max Drawdown, Sharpe Ratio, and RoMaD (Return Over Maximum Drawdown). These metrics collectively provide insights into both return and risk perspectives. For a more thorough exposition of these metrics, readers are directed to Appendix B.

\subsubsection{Backtest Results Analysis}
The results obtained from our backtesting are summarized in Table \ref{table:results}. This table elucidates how our strategy compares against the benchmark strategies across the selected metrics.

\begin{table}[ht!]
    \centering
    \begin{tabular}{|p{3cm}|p{1.75cm}|p{1.75cm}|p{1.75cm}|p{1.75cm}|}
    \hline
    Strategy & Cumulative Return & Max \mbox{Drawdown} & Sharpe \mbox{Ratio} & RoMaD \\
    \hline
    Proposed Strategy & 263\% & -20\% & 1.65 & 2.36 \\
    \hline
    Triple Barrier Labeling & 101\% & -21\% & 1.22 & 1.12 \\
    \hline
    Side Learning & 106\% & -14\% & 1.57 & 1.79 \\
    \hline
    Buy and Hold & 195\% & -55\% & 0.80 & 0.70 \\
    \hline
    \end{tabular}
    \caption{Backtest results}
    \label{table:results}
\end{table}

\begin{figure}[htbp]
    \centering
    \includegraphics[width=0.8\textwidth]{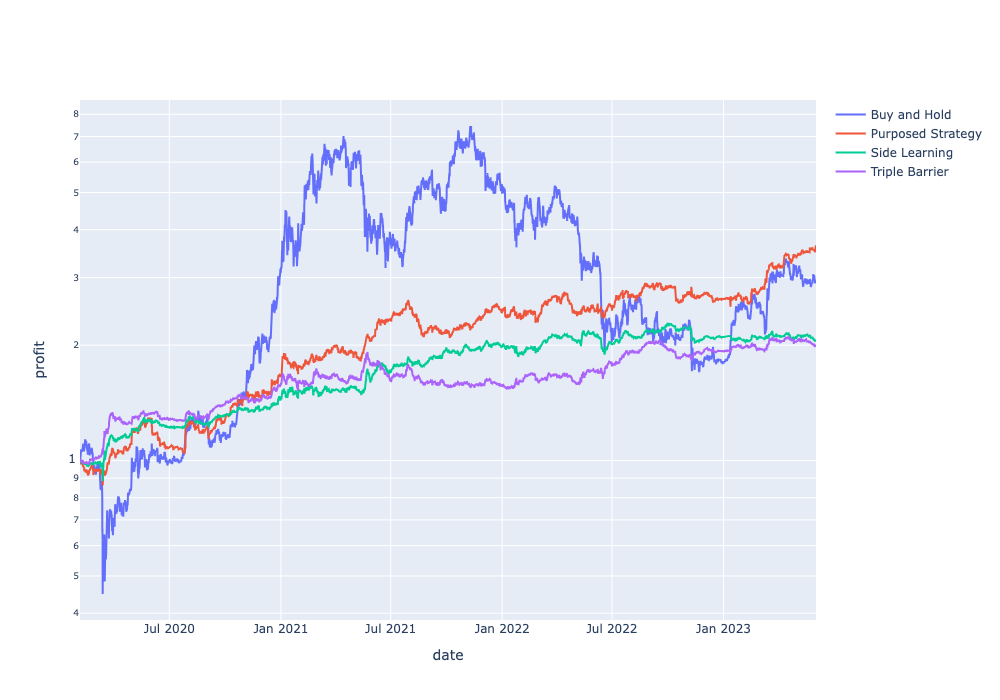}
    \caption{Backtest Results Comparision}
    \label{fig:backtest-comparision}
\end{figure}

\subsubsection{Comparative Evaluation}
As gleaned from the results, our trading strategy markedly superseded the benchmark methods across all metrics, showcasing its robustness and efficacy. Notably, it delivered superior risk-adjusted returns, as reflected in the elevated Sharpe ratio. An intriguing observation is the performance of the triple barrier labeling method. Although it aspires to capture the profit and loss potential of each trade, its logloss values imply that it couldn't match the efficacy of the regular trend prediction model, which potentially impacted its trading outcomes.

\subsection{Overall Results Summary}
In summary, our findings from both synthetic and actual prediction results echo the unparalleled benefits of a well-structured risk management strategy. These results underscore the potential of our proposed methodology in consistently delivering optimal trading outcomes, irrespective of the probabilistic or actual scenarios. The salient takeaway remains the evident enhancement in risk-adjusted returns, making our contribution pivotal in the realms of financial risk management and trading optimization.

\section{conclusion}
The application of Kelly's formula to our strategy has yielded substantial profits; however, it has also introduced a considerable level of risk. This issue can be mitigated by reducing leverage or exploring alternative risk management techniques. One such approach involves considering other formulas, such as half Kelly and others, which warrant further investigation in future research endeavors.

While we employed a straightforward model with limited optimizations, the model's accuracy could be significantly enhanced through various means. This includes incorporating more sophisticated features, exploring more technical indicators, optimizing our feature space using techniques like backward elimination, and exploring the potential of employing neural networks and other advanced models. Additionally, tuning model parameters can exert a significant influence on the overall results. An interesting avenue for improvement involves using the outcomes of the first model as features for the second model. These enhancements can be performed individually or simultaneously for both models, and their effects can be rigorously assessed.

To establish the robustness and generalizability of our findings, it is imperative to validate the results on diverse stock and currency markets. Comparing results across various time frames (e.g., minutes, hours, days) and trading volumes will provide valuable insights. Moreover, to ascertain the practical viability of our strategy, it is essential to conduct long-term trading based on the generated forecasts and compare the actual outcomes with the projected results. Our forecasts have been meticulously designed to ensure accuracy and timeliness, thereby facilitating real-time implementation in the market.

In conclusion, this study introduces a pragmatic trading approach and demonstrates its efficacy through extensive testing on the Bitcoin market. By addressing the profit-to-loss ratio and incorporating simultaneous price and direction predictions, we have achieved promising results. However, there is ample room for improvement, including risk management techniques and model optimization. Furthermore, the strategy's robustness should be examined across diverse markets and time frames. Ultimately, the actual application of our forecasts in long-term trading will serve as the litmus test for the strategy's practical value and success in the financial markets.
\section{Appendix}
\begin{appendices}
\section{Technical Indicator Formulas}
\subsection{TRX (TRIX)}
The Triple Exponential Average (TRIX) is a momentum oscillator that focuses on detecting changes in the rate of price change. It is based on triple smoothing of price and can help identify overbought and oversold conditions.
\[
TRIX = \frac{EMA(EMA(EMA(close))) - EMA(EMA(EMA(close, n), n), n)}{EMA(EMA(EMA(close, n), n), n)}
\]
where `close` is the closing price of the financial instrument for each period and `n` is the number of periods used in the exponential moving average (EMA).
\subsection{MACD (Moving Average Convergence Divergence)}
MACD is a trend-following momentum indicator that shows the relationship between two moving averages of an asset's price. It consists of the MACD line, signal line, and histogram, and is used to identify trend changes and potential buy or sell signals.
\[
MACD = EMA(close, 12) - EMA(close, 26)
\]
\subsection{PPO (Percentage Price Oscillator):}
The Percentage Price Oscillator is similar to the MACD but is displayed as a percentage rather than an absolute value. It helps traders compare the percentage difference between two moving averages.
\[
PPO = \frac{EMA(close, 12) - EMA(close, 26)}{EMA(close, 26)} \times 100
\]
\subsection{ROC (Rate of Change)}
ROC measures the percentage change in price over a specified time period. It is used to identify the strength and direction of a trend. A rising ROC indicates upward momentum, while a falling ROC suggests downward momentum.
\[
ROC = \frac{close - close_n}{close_n} \times 100
\]
Where `close` is the current closing price, and `close\_n` is the closing price `n` periods ago.

\subsection{EFI (Force Index)}
 The Efficiency Index (EFI) is a technical indicator used to measure the effectiveness of price movements in relation to trading volume.
\[
EFI = \frac{Force}{Volume}
\]
Where `Force` is the product of the price change and the volume change over a specified period and `Volume` is the trading volume for the current period.
\subsection{CMO (Chande Momentum Oscillator):}
CMO is a momentum oscillator that measures the difference between the sum of positive and negative price changes over a specified period. It helps identify overbought and oversold conditions.
\[
CMO = \frac{SMA(P) - SMA(N)}{SMA(P) + SMA(N)} \times 100
\]
Where `P` is the sum of positive price changes over a specified period, and `N` is the sum of negative price changes over the same period.
\subsection{RSI (Relative Strength Index)}
RSI is a momentum oscillator that measures the speed and change of price movements. It ranges from 0 to 100 and is commonly used to identify overbought (above 70) and oversold (below 30) conditions.
\[
RSI = 100 - \frac{100}{1 + RS}
\]
Where `RS` is the average of `n` days' up closes divided by the average of `n` days' down closes.
\subsection{CCI (Commodity Channel Index)}
CCI is a momentum-based oscillator used to identify cyclical trends in the market. It measures the deviation of an asset's price from its statistical average.
\[
CCI = \frac{1}{0.015} \times \frac{Typical Price - SMA(Typical Price, n)}{Mean Deviation}
\]
Where `Typical Price` is the average of high, low, and close prices, and `Mean Deviation` is the mean absolute deviation of the Typical Price.
\subsection{Williams \%R (Williams Percent Range)}
Williams \%R is a momentum oscillator that measures overbought and oversold conditions on a scale from -100 to 0. Readings below -80 are typically considered oversold, while readings above -20 are considered overbought.
\[
\%R = \frac{(H - C)}{(H - L)} \times (-100)
\]
Where `H` is the highest high over a specified period, `L` is the lowest low over the same period, and `C` is the most recent closing price.

\section{Metrics Explanation}

Before diving into the results of our study, it's essential to understand the key performance metrics used to evaluate and compare different trading strategies. These metrics provide insights into both the performance and risks associated with each strategy.

\subsection{Return \texorpdfstring{\( R \)}{R}}

The Return metric \( R \), often expressed as a percentage, measures the total profit or loss made from an investment over a specific period. When referencing monthly returns, it provides insight into the average growth rate of the investment on a monthly basis.

\begin{equation}
R = \frac{V_{\text{end}} - V_{\text{start}}}{V_{\text{start}}} \times 100\%
\end{equation}

Where:
\begin{itemize}
    \item \( V_{\text{end}} \) is the final value of the investment.
    \item \( V_{\text{start}} \) is the starting value of the investment.
\end{itemize}

\subsection{Drawdown \texorpdfstring{\( D \)}{D}}

Drawdown \( D \) measures the largest single drop from peak to trough in the value of a portfolio.

\begin{equation}
D = \frac{V_{\text{peak}} - V_{\text{trough}}}{V_{\text{peak}}} \times 100\%
\end{equation}

Where:
\begin{itemize}
    \item \( V_{\text{peak}} \) is the highest value achieved before a decline.
    \item \( V_{\text{trough}} \) is the lowest value reached after the peak.
\end{itemize}

\subsection{Sharpe Ratio \texorpdfstring{\( S \)}{S}}

The Sharpe Ratio \( S \), when using monthly returns, calculates how much excess return a strategy generates per unit of risk on a monthly basis.

\begin{equation}
S = \frac{R_M - R_F}{\sigma_M}
\end{equation}

Where:
\begin{itemize}
    \item \( R_M \) is the average monthly return of the trading strategy.
    \item \( R_F \) is the monthly return of a risk-free asset.
    \item \( \sigma_M \) is the standard deviation of the strategy's monthly returns, representing its risk.
\end{itemize}

\subsection{RoMaD \texorpdfstring{(\( \rho \))}{(rho)}}

RoMaD \( \rho \) gauges the risk-adjusted performance using monthly returns.

\begin{equation}
\rho = \frac{R_M}{D_{\text{max}}}
\end{equation}

Where \( D_{\text{max}} \) is the Maximum Drawdown.

By understanding these metrics thoroughly, traders and investors can make better-informed decisions when evaluating the performance and risk of different trading methodologies.

\end{appendices}

\bigskip
\noindent
{\small \textbf{Note:} During the preparation of this work, the authors used ChatGPT to improve language and readability of the article. After using this tool, the authors reviewed and edited the content as needed and take full responsibility for the content of the publication.}

\bibliography{sample}

\end{document}